\documentclass[preprint,aps]{revtex4}%
\usepackage{amsfonts}
\usepackage{graphicx}
\usepackage{amsmath}
\usepackage{amssymb}%
\begin{document}
\preprint{ }
\title{Griffiths singularities and magnetoresistive manganites}
\author{M. B. Salamon}
\email{salamon@uiuc.edu}
\affiliation{Department of Physics, University of Illinois, 1110 W. Green Street Urbana, IL 61801}
\author{S. H. Chun}
\email{schun@sejong.ac.kr}
\affiliation{Department of Physics, Sejong University, Seoul 143-747, Korea}
\date{\today}
\pacs{Numbers here}

\begin{abstract}
The large, so-called colossal, magnetoresistivity of doped manganese oxides
based on LaMnO$_{3}$ has attracted considerable attention, but only one
unusual feature of the ferromagnetic transition in these compounds. \ We
examine in this paper the progression of magnetic and thermodynamic behavior
as the transition temperature is made to vary from 360 K to 218 K by changing
the divalent dopant. \ Single crystals of La$_{0.7}$Sr$_{0.3}$MnO$_{3},$ as is
well known, show modest magnetoresistivity and conventional critical behavior.
\ La$_{0.7}$Pb$_{0.3}$MnO$_{3},$ and to an even greater extent, La$_{0.7}%
$Ca$_{0.3}$MnO$_{3},$ have unusual magnetic properties extending more than 100
K above the transition. \ We treat the properties of the latter samples in the
context of a Griffiths phase in which the transition temperature is depressed
from its maximum value $T_{G}$ by random bond-angle bending.

\end{abstract}
\maketitle

\section{\bigskip Introduction}

The properties of AMnO$_{3},$ where A is a mixture of trivalent lanthanides
and divalent ions, have intrigued researchers for decades.\cite{salamon01} The
parent compound, LaMnO$_{3}$, crystallizes in a slightly distorted perovskite
structure and is an antiferromagnetic insulator with a Ne\`{e}l temperature
$T_{N}\approx130$ K. When concentration of divalent atoms (Ca, Sr, Ba, Pb..)
substituted for La (A-site substitution) exceeds $\approx$1/8, the low
temperature phase is ferromagnetic and metallic. The Curie temperature depends
strongly on the concentration and ionic size of the substituent
\cite{hwang95b} and, perhaps most significantly, on the ionic-size variance of
A-site atoms. \cite{rodriguez96} The highest Curie temperature, $T_{C}\approx$
360 K, is achieved with Sr doping at a concentration close to 3/8; that is,
for La$_{5/8}$Sr$_{3/8}$MnO$_{3}.$ \ At this concentration, the material is
metallic in both paramagnetic ($T\geq360$ K) and ferromagnetic phases, and the
effect of magnetic fields on the electrical resistivity is not dramatic. The
ferromagnetic/paramagnetic transition is entirely normal, by which we mean
that the magnetization can be described by critical exponents very close to
those expected for a three-dimensional Heisenberg ferromagnet.
\ \cite{ghosh98}

The conventional picture for this system is based on the \emph{double
exchange} mechanism proposed by Zener.\cite{zener51} \ Each divalent
substituent converts a Mn$^{3+}$ ion to Mn$^{4+},$ with the outermost
($e_{g})$ electron on Mn$^{3+}$ site resonating with a neighboring Mn$^{4+}$
via the intervening oxygen atom. Because of strong Hund's-rule coupling, the
double-exchange transfer is favored when neighboring core spins are aligned,
leading to ferromagnetism. \ When the substitution level is sufficiently high,
the holes doped into this system form a fully spin-polarized (half-metallic)
band. As the $S=3/2$ core $(t_{2g})$ spins disorder with increasing
temperature, the resistivity increases\ and, near the Curie temperature,
exhibits substantial--though not dramatic--magnetoresistance. \ This picture
describes La$_{5/8}$Sr$_{3/8}$MnO$_{3}$ reasonably well. \cite{urushibara95}

Changing the Sr content away from La$_{5/8}$Sr$_{3/8}$MnO$_{3}$, substituting
Ca or other divalent atoms for Sr at the same concentration and even
substituting other lanthanides for La sharply decreases $T_{C}$ and
dramatically changes the nature of the paramagnetic/ferromagnetic transition.
The resistivity in the paramagnetic phase increases exponentially with
decreasing temperature, peaks somewhat above $T_{C}$, and then decreases
sharply in the ferromagnetic phase. The resistivity peak shifts to higher
temperature with increasing field, giving rise to the dramatic field dependent
resistivity that has been termed colossal magnetoresistance (CMR). A
calculation of the resistivity within the context of the double-exchange model
\cite{millis95} provided strong evidence that a localizing mechanism beyond
that model was necessary to explain these large field- and
temperature-dependent changes, and there is now strong theoretical
\cite{millis96c} and experimental evidence \cite{jaime96b} that polaron
formation and accompanying self-trapping of electrons play essential roles.
\ As the\ average ionic size of A-site atoms decreases toward that of La, the
transition temperature decreases and the exponential increase in resistivity
with temperature makes the drop to metallic resistivity at $T_{C}$ ever more
dramatic. A powerful argument can be made that the smaller the A-site atom the
greater is the distortion of the crystal from the cubic perovskite structure.
The concurrent bending of the Mn-O-Mn bond angle inhibits the double-exchange
resonance that drives ferromagnetic order and lowers $T_{C}$.
\cite{garciamunoz96}\ However, even if the \emph{average }ionic size is kept
constant (usually monitored by the so-called tolerance factor), the transition
temperature drops as the variance in ionic size increases. \cite{rodriguez96}%
\ This suggests that \emph{local} bond-angle bending is more important than
the average and that disorder therefore plays a major role. Indeed, there is
considerable evidence that metallic and polaronic regions coexist in the
vicinity of the phase transition. The phase separation is dynamic, but much
slower than is typical for critical fluctuations as can be seen in noise
measurements \cite{merithew00,podzorov00}, muon spin relaxation
\cite{heffner99}, and the presence of strong diffusive peaks in neutron
scattering. \ The case for phase separation, driven by the randomness inherent
in the system, has been documented extensively in a recent review article by
Dagotto. \cite{dagotto01}

This paper explores the dramatic changes in thermodynamic behavior that
accompany the better known changes in transport properties upon various
substitutions away from Sr$_{3/8}$. \ We will argue that bond disorder plays a
key role and that the problem should be considered in the context of a
Griffiths singularity. In his pioneering paper, Griffiths \cite{griffiths69}
considered a percolation-like problem in which each exchange bond in a system
has value $J_{1}$ with probability $p$ and $J_{2}=0$ with probability $1-p$.
\ For all $p<1$, Griffiths showed that the free energy, and thus the
magnetization, is singular at the transition point $T_{C}(p),$ a consequence
of the accumulation of clusters whose local transition temperatures exceed
$T_{C}(p).$ Fisch \cite{fisch81} extended the argument to $0\leq J_{2}<J_{1}$,
demonstrating that the singularities persist. These results suggest, as
emphasized by Dotsenko \cite{dotsenko99}, that the essential contributions of
local minima destroy the length-scaling picture of a random-fixed-point
universality class. Bray and Moore \cite{bray82} and\ Bray \cite{bray87}
extended the argument to any bond distribution that reduces the transition
temperature from some \textquotedblleft pure" value $T_{G}$ and proposed a
distribution function for the inverse susceptibility tensor that captures the
singularity proposed by Griffiths. \ Bray terms the temperature range
$T_{C}(p)\leq T\leq T_{G}$ the Griffiths phase, where $p$ is now a measure of
the bond distribution. The nature of the Griffiths singularity in the limit of
small dilution has been treated in some detail in the quantum limit where
$T_{C}(p)\rightarrow0$ by Castro Neto and
coworkers.\cite{castroneto98,castroneto00}

In this paper, which builds upon earlier work, \cite{salamon02}\ we
demonstrate the progression of the magnetic and thermodynamic properties of
doped LaMnO$_{3}$ as the transition temperature is lowered from its maximum
value. We then turn to an analysis of the low-field behavior of the
magnetization based on the eigenvalues of the inverse susceptibility as
proposed by Bray. \ In Section IV, we extend the analysis by introducing a
bond distribution that changes with temperature and field as a consequence of
the double-exchange mechanism and treat it using a cluster model. \ Section V
concludes the paper with a discussion of the implications of this analysis for
disordered double-exchange magnets.

\section{Magnetic and Thermodynamic Properties}

Three single-crystal samples were used in this study. \ Two samples,
La$_{0.7}$Sr$_{0.3}$MnO$_{3}$ (LSMO) and La$_{0.7}$Ca$_{0.3}$MnO$_{3}$ (LCMO)
were grown by optical floating-zone techniques by Okuda, et al.\cite{okuda00}
A sample of La$_{0.67}$(Pb, Ca)$_{0.33}$MnO$_{3}$ (LPMO)\ was grown by flux
methods as described elsewhere \cite{jaime98} and has a transition temperature
midway between the extremes represented by the other samples. \ The LSMO\ and
LPMO\ samples were cut into rectangular slabs with the long direction along
the $a$ directions while the LPMO was used as grown, but had a similar
orientation. \ The magnetization of each crystal was measured in a
conventional Quantum Design MPMS system with the field along the longest axis
of the sample. \ The data reported are corrected for demagnetization.
\ Following the magnetic measurements, gold current and voltage pads were
sputtered on the sample and leads were attached to the pads with silver paint.
One end of the sample was varnished to a copper block while a strain gauge
heater was attached to the opposite end. A pair of fine-wire thermocouples
were connected to measure the temperature difference between the voltage
contacts for thermopower measurements. \ The resistance and thermopower were
measured sequentially at each field-temperature point in a Quantum Design PPMS
instrument. \ Following the transport measurements, the samples were
mechanically thinned, removing the gold contact pads, and mounted for ac
calorimetry measurements. The small LPMO\ crystals were damaged in this
process and the heat capacity data in field could not be obtained. Samples
were placed in a crystat in which a magnetic field up to 7 T could be applied.
\ Light from a stabilized quartz lamp was chopped mechanically to provide
periodic heat pulses to the sample at the desired frequency. \ The proper
operating point was located at the midpoint of the range where the ensuing
temperature oscillations were inversely proportional to the frequency of the
heat pulses. \ A thorough review of the ac method has been prepared by its
inventor, Y. Kraftmakher. \cite{kraft02}

Figure 1 shows the ac heat capacity vs temperature for the three samples in
zero applied field.
\begin{figure}[tbp]
\includegraphics[width=8cm,clip]{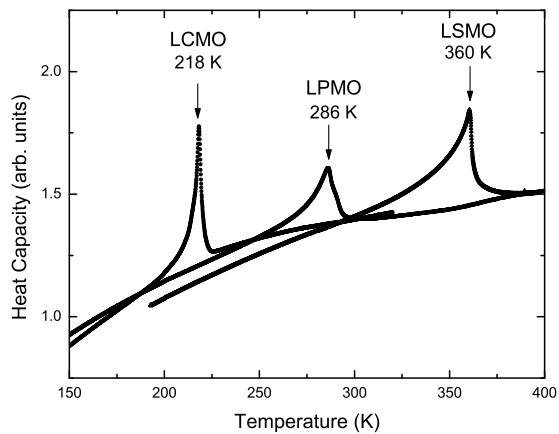}
\caption
{Heat capacity in zero applied field for the three samples. The La$_{0.7}%
$(Ca,Pb)$_{0.3}$MnO$_3$
sample was damaged upon thinning and shows a reduced
heat capacity peak.}
\label{allcp}
\end{figure}%
\ Although the LPMO\ sample shows obvious signs of the damage that accompanied
thinning, as noted above, the heat capacity exhibits a sharp peak at the
temperatures indicated as T$_{C}$ (heat capacity) in Table I. \ The heat
capacity curve for LCMO\ is significantly narrower than for LSMO, a point we
will address in more detail below. \ Despite its sharpness, there is no sign
of hysteresis in the LCMO data. \ Similarly, the magnetization curves change
significantly as the transition temperature is reduced. \ As had been reported
previously \cite{ghosh98}, the magnetization for LSMO can be collapsed to a
single curve using exponents that are similar to those expected for a
Heisenberg ferromagnet. \ Our data behave similarly, as can be seen in Fig. 2
with the exponent values given in Table I.
\begin{figure}[tbp]
\includegraphics[width=8cm,clip]{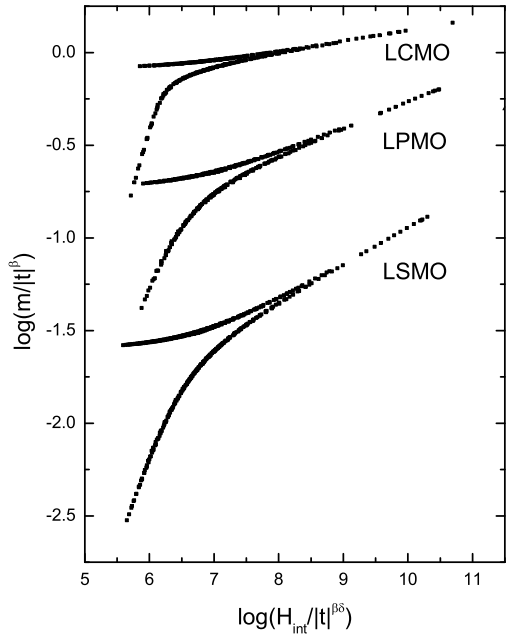}
\caption{Scaling curves for the three samples. The exponents $\beta
$ and $\delta$
deviate strongly from Heisenberg-like values as $T_C$ is reduces.  The temperatures
in parentheses indicate the range of data used in the scaling curves.}
\label{allscale}
\end{figure}%
\ Here, $t=(T/T_{C}-1)$; the values of $T_{C},$ $\alpha,$ and $\delta$ are
those that best collapse the data above (upper curve) and below (lower curve)
$T_{C}.$ The exponents are somewhat different from those reported by Ghosh et
al. \cite{ghosh98}, but are also not far from the Heisenberg values
$\beta=0.36$ and $\delta=4.8.$ \ However, as the transition temperature
decreases, the data can be collapsed only by using exponents that are far from
those for any universality class. \ %

\begin{table}[tbp] \centering
\begin{tabular}
[c]{|l|l|l|l|l|}\hline
& T$_{C}$(heat capacity) & T$_{C}$(scaling) & $\beta$ & $\delta$\\\hline
LCMO & 218 K & 216.2 K & 0.10 & 16.9\\
LPMO & 286 K & 285.1 K & 0.24 & 7.1\\
LSMO & 360 K & 359.1 K & 0.31 & 5.1\\\hline
\end{tabular}
\caption{Transition temperatures and critical exponents for samples studied.\label{TableI}}%
\end{table}
The effects can be seen more directly by following the magnetization curves
along the isotherms corresponding to the peaks in the zero-field heat capacity
curves. \ The ratio of the measured magnetization at $T_{C}$ to the
low-temperature saturation value is shown in Fig. 3 for all three samples.
\begin{figure}[tbp]
\includegraphics[width=8cm,clip]{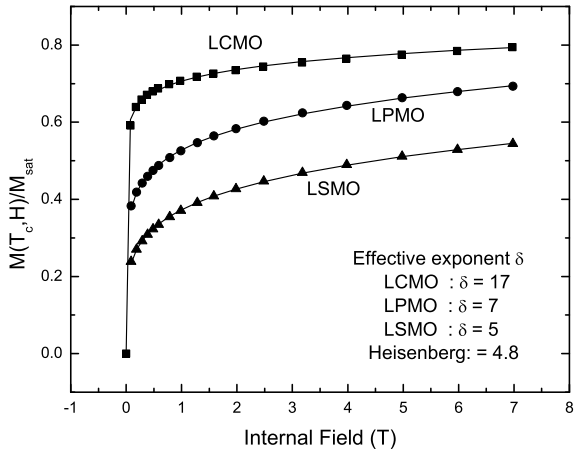}
\caption
{Magnetization vs internal field along the critical isotherm. The exponent $\delta
$
increases strongly as the transition temperature decreases.}
\label{critiso}
\end{figure}%
\ The solid curves are fits the usual expression $M(H,T_{C})\varpropto
H^{1/\delta}$ along the critical isotherm; the exponents agree with the
scaling analysis. \ Note that the magnetization of LCMO rises to 60\% of
saturation in low fields, yet shows no signs of hysteresis or remanence. \ It
is tempting to attribute this behavior to a first-order transition, but we
will discuss it in the next section in terms of a Griffiths singularity.

While the LSMO data seem quite close to Heisenberg behavior, the low field
susceptibility of that sample, as well as that of the others, is anomalous.
\ Figure 4 shows the inverse susceptibility of the three samples normalized by
the low temperature saturation magnetization $M(0)$ and plotted versus reduced
temperature $T/T_{C}.$%
\begin{figure}[tbp]
\includegraphics[width=8cm,clip]{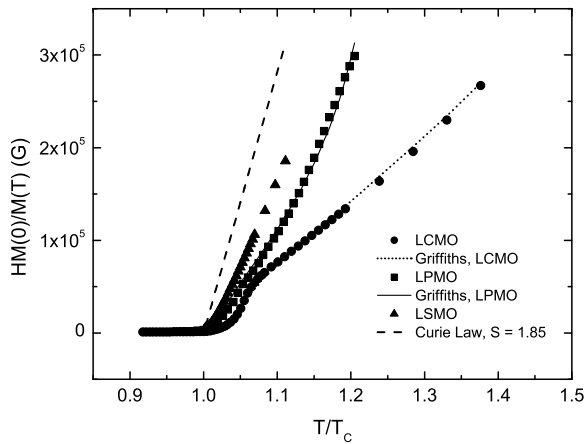}
\caption
{Inverse susceptibility multiplied by the saturation magnetization. The dashed
curve is the Curie-Weiss susceptility expected for $S=1.85$ and the critical temperature
of the LSMO sample.  The effective slope for LSMO corresponds to $S \simeq
3.5$,
increasing to $S \simeq
6$ for LCMO. The curves through the data points are for the
Griffiths model, as described in the text.}
\label{newgriff}
\end{figure}
If these data followed a Curie-Weiss law, they would lie on a straight line
given by
\begin{equation}
HM(0)/M(T)=\frac{3k_{B}T_{C}}{g\mu_{B}(S+1)}\left(  \frac{T}{T_{C}}-1\right)
.\label{Curielaw}%
\end{equation}
The dashed line is the slope expected for $T_{C}=360$ K and $S=1.85$, namely
the values for LSMO. \ The actual slope of the LSMO data corresponds to a spin
$S\approx3.5$ while that for LCMO requires $S\approx6.$ \ These results
indicate the persistance of spin clusters to temperatures significantly above
the Curie temperature, even in nominally Heisenberg-like LSMO. \ Even more
dramatic is the sharp downturn or knee in the LCMO inverse-susceptibility data
and, to a lesser but still noticeable extent, in those for LPMO. \ This
downturn, reported first by De Teresa et al. \cite{deteresa97a}, moves to
higher temperatures with increasing field. \ The scaling analysis shown in
Fig. 2 include data only for $T/T_{C}\leq1.06$; that is, at temperatures below
the downturn. We defer discussion of the other lines in Fig. 4 to the next section.

The anomalies in the magnetization are, of course, mirrored in the heat
capacity data as functions of applied field. \ Figure 5a shows the data for
LCMO and Fig. 5b, for LSMO at a succession of applied fields.
\begin{figure}[tbp]
\includegraphics[width=8cm,clip]{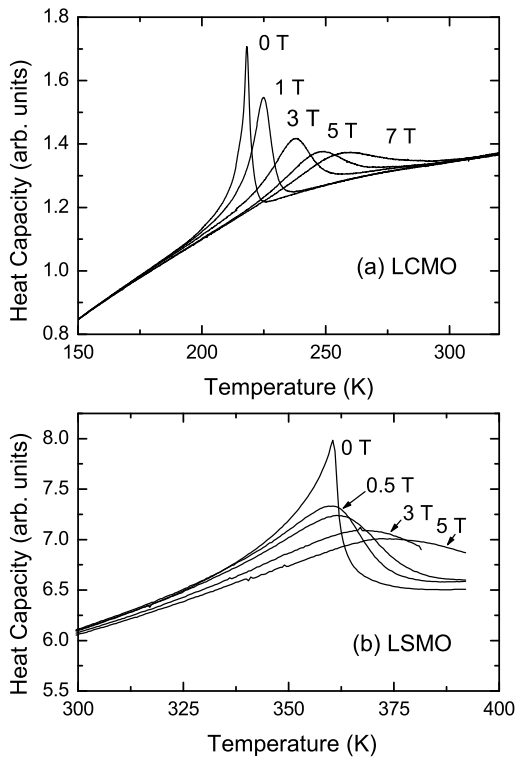}
\caption{Field dependence of the heat capacity of LCMO (a) and LSMO (b) at the
listed fields.  The curve for $B=1$ T is not labeled in b). Note the qualitatively different
behavior of the two samples. }
\label{cp_h}
\end{figure}%
\ The LCMO data shift to higher temperature while remaining relatively narrow
while the LSMO\ data, as with other ferromagnets, broaden with little shift in
peak position. \ As for the magnetization, the heat capacity data should
collapse to a universal curve when scaled with a power of the magnetic field
and plotted versus scaled temperature according to
\begin{equation}
\left(  C(H,T)-C(0,T)\right)  H^{\alpha/\beta\delta}=f(\frac{t}{H^{1/\beta
\delta}}). \label{Cpscaling}%
\end{equation}
As we reported earlier,\cite{lin00} neither the exponents that provide a
scaling collapse of the magnetization data, nor any other set that we can
identify, are able to satisfy the scaling conditions for LCMO. \ This is shown
in Fig. 6a.
\begin{figure}[tbp]
\includegraphics[width=8cm,clip]{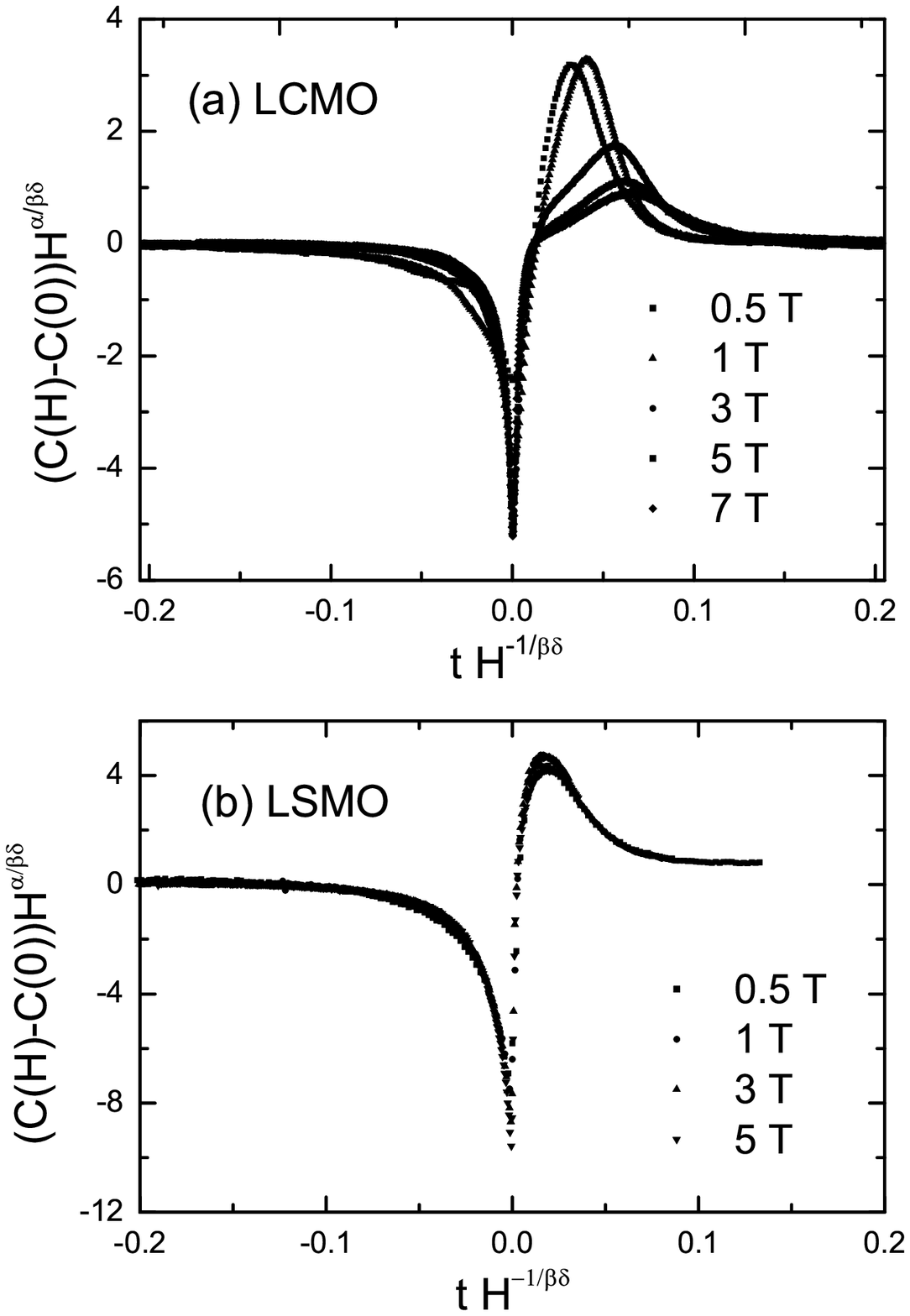}
\caption
{The deviation of the heat capacity in magnetic field from its zero-field value,
scaled by field and plotted versus scaled reduced temperature. No set of exponents \
can be found to collapse the LCMO data in a) to a single curve.  By contrast, the same
values of $\beta$ and $\delta$ used in Fig. 2, along with $\alpha
= -0.1$ serve to
collapse the LSMO data in b).}
\label{cpscale}
\end{figure}%
\ However, the LSMO\ data, Fig. 6b, do fall on a single scaling curve using
the values of $\beta$ and $\delta$ from the magnetization scaling, and
$\alpha=-0.1;$ the last differs slightly from a value consistent with $\beta$
and $\delta.$ As the susceptibility data of Fig. 4 demonstrate, even LSMO does
not exhibit single-spin behavior, so we must take the critical exponents to
represent only effective values.

\section{Griffiths Phase Analysis: Susceptiblity}

In his pioneering 1969 paper, Griffiths \cite{griffiths69} demonstrated that
the magnetization of a randomly diluted ferromagnet above its percolation
point is a non-analytic function of the field at all temperature below the
pure-system Curie temperature. \ The argument was extended to alloys; i.e.,
for $0\leq J_{2}<J_{1},$by Fisch \cite{fisch81} and to any positive-definite
(bounded) distribution of exchange interactions by Bray and Moore.
\cite{bray82} \ In the latter paper, the authors focused on the distribution
$\rho(\lambda)$ of the eigenvalues $\lambda$ of the inverse susceptibility
matrix. Above the critical temperature $T_{C}$ but below the highest
achievable critical temperature $T_{G}$, all states with small values of
$\lambda$ are localized; there are local regions of large susceptibility, but
no long range order. Just at $T_{C},$ an extended state of infinite
susceptibility $(\lambda=0)$ appears, signalling the sudden onset of
long-range order. Subsequently, Bray \cite{bray87} suggested an explict form
for this distribution,%
\begin{equation}
\rho(\lambda)\propto\lambda^{-x}\exp(-A(T)/\lambda).\label{rho(lambda)}%
\end{equation}
The power-law prefactor was not specified but Bray and Huifang later
\cite{bray89} considered a soluble model of the diluted Ising ferromagnet and
verified Eq. (\ref{rho(lambda)}) with $x=1/2.$ The amplitude $A$ was argued to
diverge as $(1-T/T_{G})^{-2\beta}$at the pure, or Griffiths, temperature
$T_{G}$ and to vanish as $\left(  T/T_{C}-1\right)  ^{2(1-\beta)}$ at the
actual Curie point.\ The exponent $\beta$ is the usual exponent for the system
at its pure transition$.$This distribution peaks at $\lambda=A/x$ and vanishes
at $\lambda=0$ for all temperatures above $T_{C}.$ There is, therefore, a
pile-up of small eigenvalues--large susceptibilities--as the Curie temperature
is approached. Just at $T_{C}$ the distribution collapses into $\lambda=0$
causing the magnetization to jump to a large value in applied field--the
hallmark of the Griffiths singularity.

We assert here that the transition temperatures evidenced in the sequence
LSMO, LPMO and LCMO are a consequence of increased randomness due to the
increased local bond bending in the vicinity of successively smaller dopant
atoms. \ If so, then each sample is farther below the Griffiths temperature of
an optimal system and will consequently exhibit a broader temperature range
over which $A(T)$ varies between its zero at $T_{C}$ and its divergence at
$T_{G}$. \ We calculate the average susceptibility from Eq. (\ref{rho(lambda)}%
) according to%
\begin{equation}
\overline{\chi}=C\frac{\int_{0}^{T}\lambda^{-1}\rho(\lambda)d\lambda}{\int
_{0}^{T}\rho(\lambda)d\lambda}, \label{avchi}%
\end{equation}
where $C=ng^{2}\mu_{B}^{2}S(S+1)/3k_{B}$ is the Curie constant and the upper
limit of the integral recognizes that the smallest susceptibility at any
temperature is $C/T$ for spin $S.$ The exponential amplitude is taken to be%
\begin{equation}
A(T)=a\frac{(T/T_{C}-1)^{2(1-\beta)}}{(1-T/T_{G})^{2\beta}}, \label{A(T)}%
\end{equation}
with $\beta=0.38$ and $a,T_{C},T_{G}$ and $x$ varied to fit the susceptibility
data. \ The down-turn in the inverse susceptibility curves sets the value of
$T_{C}$ while the upward curvature is controled by $T_{G.}$ There is
considerable covarience of the amplitude $a$ and prefactor expononent $x,$ so
the values are subject to some uncertainty. We use the effective spin $S=1.85$
appropriate for 70\% $S=2$ and 30\% $S=1.5.$ Because the downturn (if there is
one) for LSMO is not discernable, we cannot get an unambiguous fit for those
data. \ However, the solid curves for LPMO and LCMO are reliable, with the
parameter values given in Table II.%

\begin{table}[tbp] \centering
\begin{tabular}
[c]{|l|l|l|l|l|}\hline
& $a$ (K) & $x$ & $T_{C}$ (K) & $T_{G}$ (K)\\\hline
LCMO & 5.0 & 0.53 & 224.8 & 376\\
LPMO & 4.15 & 0.61 & 293.5 & 365\\\hline
\end{tabular}
\caption{Parameters used in Griffiths susceptibility calculation.\label{key}}%
\end{table}%

Of considerable interest is the fact that the Griffiths temperatures that
emerge from the fits are comparable and only slightly above the observed
$T_{C}$ for LSMO. \ This indicates that LSMO\ lies very close to the optimal
critical temperature and explains why it can be treated in the context of an
ordinary Heisenberg ferromagnet, albeit with slightly modified critical exponents.

Note that the critical temperature obtained from the Griffiths fit is somewhat
higher than that obtained from scaling or the heat capacity. This may reflect
the suggestion made by Griffith in his original paper that the susceptibility
would tend to diverge in advance of the onset of long-range order. \ To
examine this, we focus on the downturn in the inverse susceptibility for LCMO.
\ In recent work on f-electron compounds in which disorder has driven $T_{C}$
to 0 K, Castro Neto, et al. \cite{castroneto98} have argued that the
susceptibility diverges as $T^{y-1}$ where $y\leq1$ is related to the
tunneling barrier for a cluster of $N$ \ aligned spins. In general, the
relaxation rate of Griffiths clusters is also expected to be proportional to
its inverse susceptibility \cite{bray87}, so similar arguments might hold
here; i.e. $\chi^{-1}\varpropto(T/T_{C}-1)^{1-y}.$ In Fig. 7
\begin{figure}[tbp]
\includegraphics[width=8cm,clip]{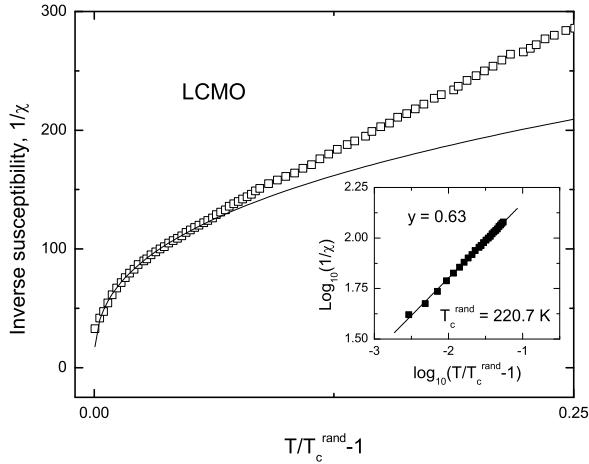}
\caption
{Inverse susceptibility of LCMO at 500 Oe as a function of $T/T_c$.  The solid curve
in the main figure and the logarithmic plot in the inset is a fit of the data
to a power law with the result that $y=0.63$.}
\label{Fig7}
\end{figure}
we plot the low-field susceptbility of the LCMO\ crystal as a function of
$T/T_{C}$, with $T_{C}=220.7$ K obtained by fitting the data to a power law.
\ The random critical temperature is much closer to that indicated by the heat
capacity peak, and is a more reliable measure of the tendency of the inverse
susceptibility to vanish with an exponent $y=0.63;$ that is, to approach
$T_{C}$ with infinite slope. Though closer to the heat capacity peak (218.2 K
at this field) it appears to be somewhat above the temperature at which
long-range order is established, as suggested by Griffiths.

\section{Griffiths Phase Analysis:\ Heat Capacity}

In the classic Griffiths-phase model, exchange interactions are distributed
randomly, but once distributed, are fixed. \ This is not the case for a
double-exchange system in which the effective coupling between two Mn ions
depends on the alignment of their respective core spins or, equivalently, the
rate at which the outer $e_{g}$ electron hops between the two ions. \ As a
consequence, as spins order locally, the spin clusters are also more metallic,
and the combined effect is to reinforce and stabilize the formation of large,
Griffiths clusters. In the presence of an applied magnetic field, these
metallic, spin-aligned clusters form at higher temperatures, strongly
affecting the thermodynamics of the transition and, of course, giving rise to
the CMR\ effect itself.

The heat capacity associated with the Griffiths singularity was studied for
the random spherical model by Rauh, \cite{rauh76} who found a jump singularity
at $T_{C}.$ We take a different approach here, using the Oguchi model
\cite{smart} to calculate the magnetization and the associated short-range
order parameter. \ In this approach, the interaction energy of a pair is
calculated exactly using the double exchange energy%
\begin{equation}
E_{de}(S_{t})=-xt\frac{S_{t}+1/2}{2S+1}-\overline{E_{de}(S_{t})},\label{Ede}%
\end{equation}
where the bar denotes an average over all possible values of the total spin
$1/2\leq S_{t}\leq7/2$ of the two $S=3/2$ cores and the shared $e_{g}$
electron of the pair. The pair interacts with its $z-1$ neighbors through an
effective magnetic field%
\begin{equation}
H_{eff}(H,T)=H+2(z-1)S\left\{  c(H,t)J_{met}+\left[  1-c(H,T)\right]
J_{ins}\right\}  m(H,T).\label{Heff}%
\end{equation}
Here, $m(H,T)$ is the reduced magnetization to be calculated, $J_{met}$ is the
exchange intereraction in metallic regions that have a concentration $c(H,T),$
and $J_{ins}$ is the exchange energy in non-metallic (but still conductive)
regions. The insulating exchange energy can be extracted directly from the
inverse susceptibility by extrapolating the linear region of Fig.
\ref{newgriff} to obtain the Curie temperature $\Theta=202$ K, from which
mean-field theory gives $J_{ins}=0.85$ meV. We obtain $J_{met}$ from the
spin-wave dispersion of manganites which is $D\approx160$ meV \AA $^{2}$
independent of concentration. The effective Heisenberg exchange interaction
giving this spin-wave stiffness is $J_{met}=D/2S_{eff}a^{2}=1.56$ meV; here
$S_{eff}=1.85$ is the average spin per manganese atom. The hopping energy
giving the same spin-wave spectrum is $t=$ $D(2S+1)/xa^{2}=$ 140 meV.
\cite{ohata73} Alternatively, the critical temperature from Monte Carlo
simulations is $k_{B}T_{C}\approx0.14t$ \cite{calderon98}, giving $t=134$ meV.

The most important input into the model is the relative concentration of
metallic bonds. We obtain this empirically from the resisitivity data as
outlined in Fig.8.
\begin{figure}[tbp]
\includegraphics[width=8cm,clip]{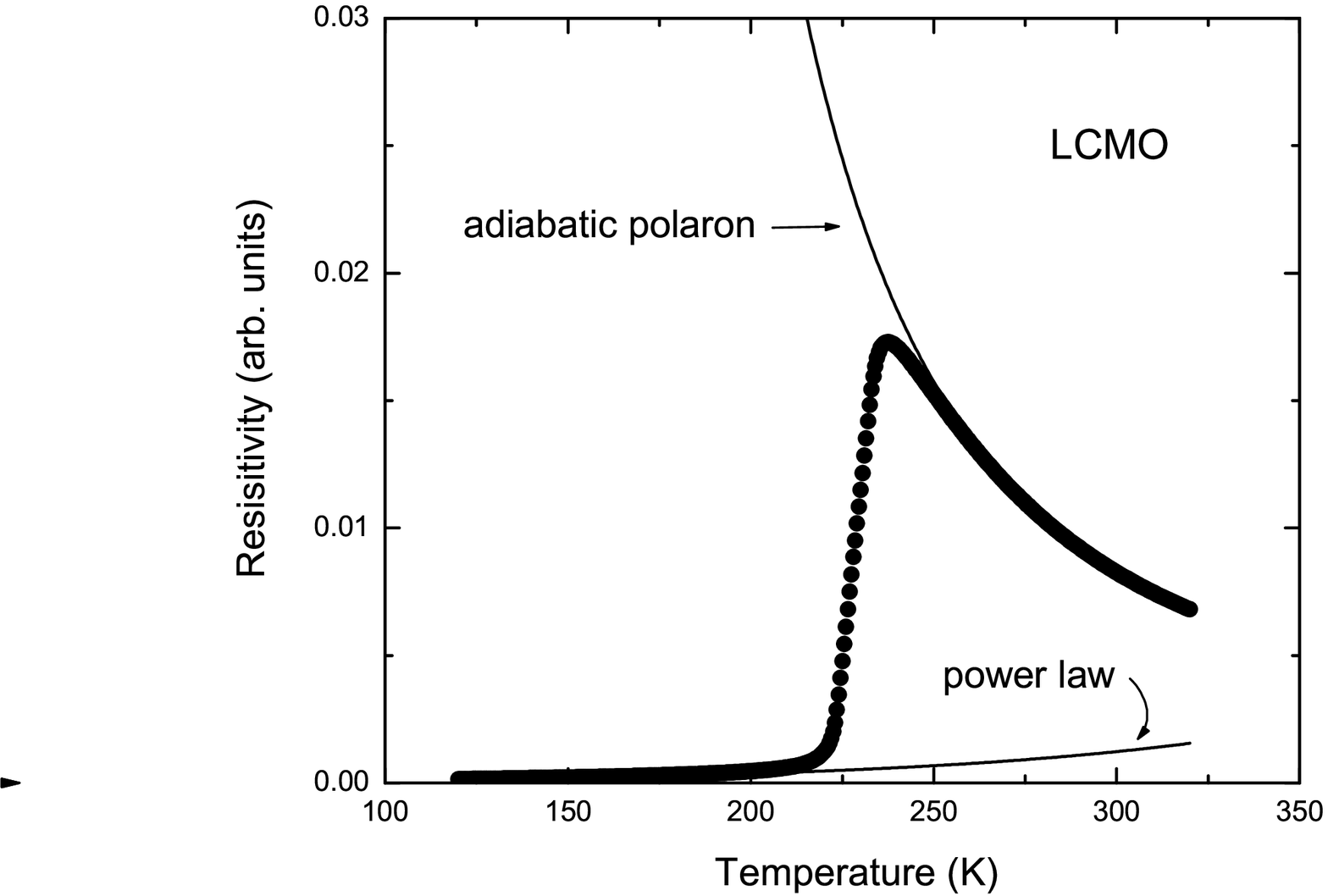}
\caption
{Resistivity of the LCMO sample in a field of 1T. Superposed are the fits to the
zero-field data at low tempratures to a power law and at high temperatures, to an
adiabatic small polaron model.}
\label{Fig8}
\end{figure}
\ The zero-field data at low temperature are fit to the power law
\begin{equation}
\rho_{lt}=\rho_{0}+a_{2}T^{2}+a_{5}T^{5},\label{rhomet}%
\end{equation}
and the high-temperature data, to an adiabatic small-polaron contribution,%
\begin{equation}
\rho_{ht}=bT\exp(E_{p}/T),\label{rhopol}%
\end{equation}
as done previously. \cite{jaime99a} The metallic fraction is obtained by
solving the generalized effective medium (GEM) expression \cite{mclachlan93}
using the experimental resistivity $\rho_{\exp}$ and the extrapolated high and
low temperature fits. \ The GEM approach guarantees that percolation occurs at
a critical concentration $c_{c}$ which we set to the 3D value for spherical
inclusions, namely $c_{c}\approx1/6.$ \ The equation to be solved for $c(H,T)$
is%
\begin{equation}
c(H,T)\frac{\rho_{\exp}^{1/t}-\rho_{lt}^{1/t}}{\rho_{\exp}^{1/t}+A\rho
_{lt}^{1/t}}+\left[  1-c(H,T)\right]  \frac{\rho_{\exp}^{1/t}-\rho_{ht}^{1/t}%
}{\rho_{\exp}^{1/t}+A\rho_{ht}^{1/t}}=0,\label{GEM}%
\end{equation}
where $A=(1-c_{c})/c_{c}$ and the percolation exponent is set to $t=2.$
Several resistivity curves and the resulting concentrations are shown in Fig.
9.
\begin{figure}[tbp]
\includegraphics[width=8cm,clip]{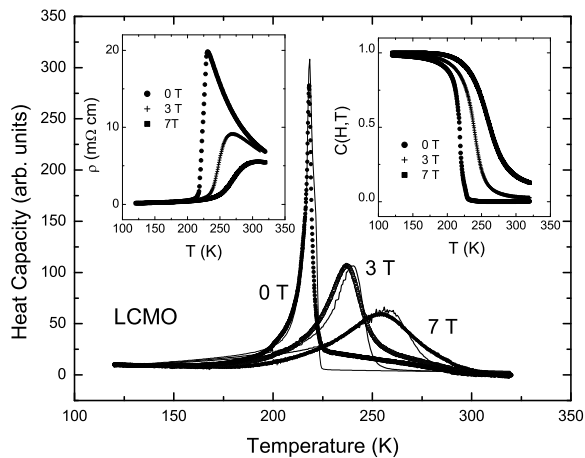}
\caption
{Heat capacity curves calculated from the Oguchi model at several fields.
The metallic concentrations extracted from the resistivity curves (left inset) are shown
in the right inset.  The only other input is the overall amplitude of zero-field curve.}
\label{Fig9}
\end{figure}%

We proceed by calculating the magnetization self-consistently in the context
of the Oguchi model; that is, we solve%
\begin{equation}
m(H,T)=\frac{1}{Z}%
{\displaystyle\sum\limits_{S_{t}=1/2}^{7/2}}
\sum_{p=S_{t}}^{S_{t}}p\exp\left(  \frac{-E_{de}+pg\mu_{B}H_{eff}(H,T)}%
{k_{B}T}\right)  , \label{m-oguchi}%
\end{equation}
where $Z$ is the partition function (same sum without the factor $p$ ). Once
$m(H,T)$ is known, we compute the energy density by averaging $E_{de}(S_{t})$
at each field/temperature point using the Boltzmann factors that have been
calculated self-consistently, and differentiate numerically to obtain the heat
capacity. \ The amplitude is chosen to fit the zero-field data and kept
constant for other fields. \ The results are shown in Fig. \ref{Fig9}. \ The
width, amplitude, and shift in temperature of curves at successive fields
agrees extremely well with the data. \ In each case, the experimental peaks
are broader on the high-temperature side of the curve, indicating that the
Oguchi calculation underestimates the persistence of short-range order to
higher temperatures. Nonetheless, the curves show quite clearly that the
metallic concentrations extracted from the GEM analysis are able to predict
the unusual critical behavior of LCMO.

The final question in this analysis is whether the Oguchi model described here
actually reproduces Griffiths-like behavior at low fields. \ The magnetization
has been calculated at the same fields as the data in Fig.\ref{newgriff},
using the zero-field value $c(0,T)$ extracted from the resistivity. \ The
result is shown in Fig.10.%
\begin{figure}[tbp]
\includegraphics[width=8cm,clip]{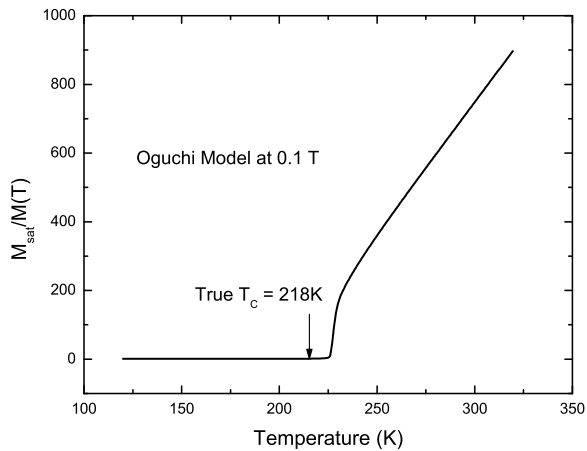}
\caption
{Inverse susceptibility at low field calculated using the Oguchi model. The sharp
downturn at a temperature above the peak in the heat capacity peak mirrors the
behavior of the experimental data.}
\label{Fig10}
\end{figure}
\ Note that the down-turn in advance of the heat capacity peak is similar to
the experimental data. \ The Oguchi approach does not capture the persistence
of spin clusters to the Griffiths temperature.

\section{Conclusions}

The nature of the ferromagnetic-paramagnetic transition in these materials has
been subject to considerable discussion. \ Most recently, Kim, et
al.\cite{kim02}argued for a tricritical point just above $x=0.33$ in
La$_{1-x}$Ca$_{x}$MnO$_{3}$. \ They do not identify the two phase lines that
emanate from the tricritical point. \ Further, examination of the $x=0.33$
data shows that the susceptibility at 290 K corresponds to a spin $S=3.5,$
rather than the $S\leq2$ expected. \ Consequently, even 30 K above the
transition there is evidence for clustering, a signature of the Griffiths
behavior we propose here. \ Apart from the critical behavior, various
explanations for the CMR\ phenomenon drawn upon elements of the
Griffiths-phase approach--mixed phases and phase separation, percolation, slow
dynamics--but have not connected them into a coherent picture. In particular,
the dramatic changes in behavior that accompany subtle changes the size and
concentration of dopant atoms have not been adequately treated. \ We have
attempted here to demonstrate that the intrinsic randomness introduced by
substituting ions that differ in size (and of course valence) from the usual
A-site atom drive the system from its optimal doping and ionic size at
Sr$_{3/8}$ to the strong CMR regime as Sr is changed to Pb and finally Ca.
\ Remarkably, the transition to the magnetic phase remains second-order like,
by which we mean that the properties are fully reversible and, with the
exception of the heat capacity, can be treated by the usual ferromagnetic
scaling equations, albeit with non-universal (even bizarre) values for the
critical exponents.

Outside the "critical" regime, there is ample evidence in our data, and in a
wealth of further data in the literature, to demonstrate coexistence of more
or less metallic and more or less insulating regions over a wide temperature
range both above and below the Curie temperature. \ We have shown that the
clusters evolve as the temperature is reduced toward $T_{C}$ in a manner
consistent with the theoretical ideas of Bray \ and Moore \cite{bray82} and
Bray \cite{bray87}. In essence, the transition is not primarily a question of
connectedness and the evolution of a tenuous infinite cluster, but rather more
a homogeneous nucleation problem in which the most-probable cluster size grows
as the temperature is reduced until they become effectively space-filling,
providing an abrupt onset of nearly complete long-range order.

The situation in the manganites differs significantly from straightforward
Griffiths phase precisely because Griffiths clusters are more metallic and
therefore more ferromagnetic than the surrounding matrix. \ the CMR effect
thus reinforces cluster formation: local spin ordering increases the mobility
of electrons, which then increases local exchange interactions via double
exchange, which in turn feeds back to lock local spin ordering. \ We have
attempted to deal with this effect phenomenologically by determining the
fraction of metallic, high-susceptibility clusters from the field and
temperature dependent resistivity using a generalized effective medium
approach. \ Knowing that fraction, we compute an effective magnetic field
acting on each pair of double-exchange coupled spins and from that, determine
the magnetization and energy density. \ We demonstrate that this approach
accurately tracks the height and temperature of the peak in the heat capacity
and, to a significant extent, its width. \ We regard the unusual behavior of
the heat capacity in magnetic field, along with the strongly non-Curie-Weiss
behavior of the susceptiblity to be hallmarks of the CMR effect, as important
in understanding it as the more dramatic changes in transport property.

Our analysis of the CMR transition in terms of Griffiths-phase ideas provides
an understanding of the evolution of behavior from LSMO, whose Curie point is
near the Griffiths temperature, to LCMO, which exhibits Griffiths phase and
magnetotransport signatures. However, the interplay of local order and
enhanced double exchange requires empirical input and remains, therefore,
unsatisfactory. \ We still need to understand the mechanism by which Griffiths
clusters order in a polaronic, double-exhange magnet, and how that process
assists in stabilizing large clusters. \ It is our hope that this paper has
helped to delineate the problems that remain.

\section{Acknowledgements}

This material is based upon work supported by the U.S. Department of Energy,
Division of Materials Sciences under Award No. DEFG02-91ER45439, through the
Frederick Seitz Materials Research Laboratory at the University of Illinois at
Urbana-Champaign. We are especially grateful for the crystals provided by T.
Okuda, Y. Tomioka, and A. Asamitsu in the group of Prof. Y. Tokura. We have
also benefited from discussions with Marcelo Jaime, Yuli Lyanda-Geller, and
Paul Goldbart.

\bibliographystyle{apsrev}
\bibliography{BIBCMR,CrO2paper}
\ 
\end{document}